\documentclass[a4paper,11pt]{article}

\usepackage{jcappub} 
\usepackage[utf8]{inputenc}
\usepackage[T1]{fontenc}
\DeclareUnicodeCharacter{202F}{\,}

\usepackage{array}
\usepackage{makecell} 

\usepackage{caption}

\usepackage[mathscr]{eucal}
\usepackage[Symbolsmallscale]{upgreek}
\usepackage{xcolor}
\usepackage{slashed}
\usepackage{hyperref}
\usepackage{tikz}
\usepackage{subcaption}
\usepackage{tikz-feynman}
\usepackage{cancel}
\usepackage{amsmath}
\usepackage{bm}

\definecolor{darkgreen}{rgb}{0.0,0.5,0.0}

\title{Supernova-Boosted Dark Matter \\
at Large-Volume Neutrino Detectors}

\author[1]{Badal Bhalla,}
\author[1]{Fazlollah Hajkarim,}
\author[2]{Doojin Kim,}
\author[1]{and Kuver Sinha}

\affiliation[1]{Homer L. Dodge Department of Physics and Astronomy, \\ University of Oklahoma, Norman, OK 73019, USA}
\affiliation[2]{Department of Physics, University of South Dakota, Vermillion, SD 57069, USA }

\emailAdd{badalbhalla@ou.edu}
\emailAdd{fazlollah.hajkarim@ou.edu}
\emailAdd{doojin.kim@usd.edu}
\emailAdd{kuver.sinha@ou.edu}

\abstract{
Core-collapse supernovae, among the universe's most energetic events, offer a novel window into the dark sector by potentially producing a flux of boosted dark matter (BDM). We explore the potential to detect the BDM produced by supernovae with a focus on fermionic dark matter that interacts with the visible sector through a dark gauge boson. We consider the expected BDM flux at Earth, originating from both the diffuse background of all galactic supernovae and potentially strong signals from individual nearby events. Focusing on BDM-electron scattering, we project the sensitivity of major current and future large-volume neutrino detectors---DUNE, Hyper-Kamiokande, and JUNO---to this elusive signal. Our results indicate that these experiments can significantly constrain or discover BDM within compelling parameter spaces, with sensitivity notably enhanced during nearby supernova occurrences. We further emphasize the unique multi-messenger opportunity presented by a galactic supernova, where the characteristic time delay between the neutrino burst and the BDM signal arrival could provide powerful evidence and enable probes of dark matter properties.
}

\begin{document}

\maketitle
\flushbottom

\section{Introduction}
\label{sec:intro}


Over the past few decades, the quest to unravel the fundamental nature of dark matter has stood at the forefront of modern physics, bringing together particle physicists, astrophysicists, and cosmologists in a concerted effort to explore the enigmatic constituents of the Universe~\cite{Bertone:2004pz,Feng:2010gw,Olive:2003iq}. Despite the extraordinary precision of observational cosmology, which demonstrates that dark matter accounts for a significant fraction of the total energy budget of the Universe~\cite{Aghanim:2018eyx,Hinshaw:2012aka}, its underlying properties---such as mass, spin, and interaction cross sections---remain elusive. Weakly interacting massive particles (WIMPs)~\cite{Jungman:1995df,Steigman:1984ac,Bergstrom:2000pn} have been one of the most widely studied dark matter candidates that are probed by various direct, indirect, and collider experiments~\cite{Aprile:2018dbl,Akerib:2019fml,ATLAS:2017nga}. However, no conclusive evidence has yet emerged for the canonical WIMP model, prompting novel theoretical ideas and experimental approaches. Among these emerging ideas is the concept of \textit{boosted dark matter} (BDM)~\cite{DEramo:2010keq,Huang:2013xfa,Agashe:2014yua,Berger:2014sqa,Kong:2014mia,Bhattacharya:2014yha, Kopp:2015bfa, Cherry:2015oca,Necib:2016aez,Alhazmi:2016qcs,Kim:2016zjx,Bhattacharya:2016tma,Cui:2017ytb,Giudice:2017zke,Chatterjee:2018mej, Kim:2018veo, Aoki:2018gjf, McKeen:2018pbb, Kim:2019had, Berger:2019ttc, Heurtier:2019rkz, Kim:2020ipj, DeRoeck:2020ntj, Fornal:2020npv, Alhazmi:2020fju, Chigusa:2020bgq, Borah:2021jzu, Borah:2021yek, Toma:2021vlw,Bringmann:2018cvk, Ema:2018bih, Cappiello:2019qsw, Dent:2019krz, Wang:2019jtk, Ge:2020yuf, Cao:2020bwd, Jho:2020sku, Cho:2020mnc, Dent:2020syp, Bell:2021xff, Xia:2021vbz,Jho:2021rmn, Das:2021lcr, Chao:2021orr,Kouvaris:2015nsa, Hu:2016xas, An:2017ojc, Emken:2017hnp, Calabrese:2021src, Wang:2021jic,Alvey:2019zaa, Su:2020zny, Acevedo_2024}, wherein dark matter particles can receive a significant velocity or energy boost through various astrophysical processes in the present universe, thus leaving relativistic signatures of such dark matter in terrestrial detectors. Inspired by this novel idea, various neutrino and dark-matter direct detection experiments have conducted searches for BDM~\cite{Super-Kamiokande:2017dch,Ha:2018obm,PandaX-II:2021kai,Super-Kamiokande:2022ncz,PandaX:2024pme}.

In parallel, neutrino detectors have become increasingly sophisticated and play a crucial role in multi-messenger astrophysics~\cite{Giunti:2007ry, Zuber:2020kci, Casper:2002sd}. Originally developed to study neutrino oscillations and measure low-energy solar neutrinos, large-scale underground and underwater/ice-based detectors---such as Super-Kamiokande (SK)~\cite{Fukuda:1998mi,Super-Kamiokande:2010tar,Super-Kamiokande:2014ndf,Super-Kamiokande:2015qek}, SNO~\cite{SNO:2001kpb,Andringa:2015tza}, Borexino~\cite{Alimonti:2008gc,Borexino:2012udf}, KamLAND~\cite{Abe:2008aa,KamLAND:2014gul,KamLAND:2015dbn}, IceCube~\cite{IceCube:2013dkx,IceCube:2013low,IceCube:2016zyt}, and JUNO~\cite{JUNO:2015zny,JUNO:2015sjr,Guo:2019vuk}---have evolved into versatile instruments capable of detecting cosmogenic signals from a variety of physics channels, including supernova neutrinos~\cite{Krauss:1983zn}, atmospheric neutrinos~\cite{Fogli:2012ua}, and potential signatures of exotic physics beyond the Standard Model~\cite{Horiuchi:2013noa}.  
Down the line, near-future neutrino detectors, including DUNE~\cite{DUNE:2018hrq,DUNE:2018mlo,DUNE:2020lwj,DUNE:2020ypp,DUNE:2020txw} and Hyper-Kamiokande (HK)~\cite{Abe:2011ts,Hyper-Kamiokande:2016srs,Hyper-Kamiokande:2018ofw}, are expected to advance this multi-faceted physics program with greater precision.
The unique capabilities of these neutrino detectors, such as exquisite sensitivity to energy depositions as low as $\sim {\rm MeV}$ and extended exposure times that compensate for (relatively) low signal flux, have enabled their use as {\it dark matter detectors} in scenarios where BDM induces relativistic signatures in the present universe.

Recent theoretical developments emphasize that core-collapse supernovae may play a decisive role in dark matter phenomenology~\cite{Bethe:1990mw,Burrows:2012ew,peebles1984dark}. The high densities and extreme energies characteristic of supernova interiors create a fertile environment for particle production and scattering processes that can ``boost'' dark matter particles to energies well above their average halo velocities~\cite{Ghosh:2024dqw,Blanco:2019hah}.  \textit{Dark matter from supernovae} has thus emerged as a novel avenue to explore the properties of dark matter, particularly by exploiting the synergy between BDM searches and neutrino detectors~\cite{Murase:2008sp,Blum:2016afe,Suliga:2020vpz}.  In this picture, a supernova event can serve as an intense source of BDM, endowing the particles with energies well beyond typical galactic dark matter velocities and yielding observable signals in detectors originally commissioned for neutrino physics. Moreover, the typical timing of supernova neutrino bursts, combined with any directionality information offered by certain detector technologies, can help discriminate potential BDM-induced events from backgrounds~\cite{Blum:2016afe,Suliga:2020vpz}.

While the prospect of detecting BDM from supernovae is theoretically compelling, addressing uncertainties in supernova modeling---particularly regarding dark matter production mechanisms and the nuclear equation of state---is crucial for a more robust understanding~\cite{Fischer:2017lag}. Variations in progenitor mass, metallicity, and rotation can alter the energy budget and particle flux. Furthermore, the strength of dark matter interactions with nucleons or leptons, along with the mass of any new mediator, significantly influences the likelihood of dark-matter production and escape~\cite{Raffelt:1990yz,Ghosh:2024dqw}. Advancing supernova simulations and improving nuclear physics constraints will be essential for refining flux estimates and enhancing the discovery potential of BDM in neutrino detectors.

In spite of these uncertainties, investigating supernova-induced BDM search prospects remains highly compelling, as successful detection could provide unique insights into both dark matter properties and core-collapse supernova dynamics. The potential to probe new physics in extreme astrophysical environments and complement terrestrial dark matter searches makes this avenue of exploration particularly valuable. The beginning effort in this direction was made in Ref.~\cite{DeRocco:2019jti} within the context of ton-scale dark matter direct detection experiments. Diffuse fast-moving or boosted MeV-scale dark matter produced by supernova explosions carries sufficient energy to generate detectable signals through its interactions with nuclei, offering a pathway to explore previously uncharted regions of parameter space. 

While nuclear scattering serves as a viable detection channel, electron scattering is expected to provide complementary insights in the search for supernova-induced BDM, as a significant fraction of the incoming BDM energy will be transferred to recoil electrons. However, typical ton-scale dark matter direct detection experiments are optimized for keV-to-sub-MeV energy deposits, whereas electron recoils from such BDM can extend well beyond the MeV range. These large energy deposits may lead to detector saturation issues, necessitating dedicated event selection strategies to enhance sensitivity to general BDM signals~\cite{Giudice:2017zke}. Nevertheless, access to the relevant parameter space is limited, and large-volume neutrino detectors play a crucial role, further benefiting from their substantial detection mass, which compensates for the low signal flux~\cite{Kim:2020ipj}. 

In this context, we explore the detection prospects of supernova-induced BDM signals at large-volume neutrino detectors such as DUNE~\cite{DUNE:2018hrq,DUNE:2018mlo,DUNE:2020lwj,DUNE:2020ypp,DUNE:2020txw}, HK~\cite{Abe:2011ts,Hyper-Kamiokande:2016srs,Hyper-Kamiokande:2018ofw}, and JUNO~\cite{JUNO:2015zny,JUNO:2015sjr,Guo:2019vuk}. Using the conventional vector-portal dark matter scenario as a benchmark, we first analyze the prospects for detecting diffuse supernova-induced BDM signals. Additionally, we examine the sensitivity to BDM from a nearby supernova explosion occurring in the near future, taking Betelgeuse as an illustrative example. Notably, such BDM signals from local sources contribute to the emerging field of multi-messenger astrophysics, which integrates electromagnetic, neutrino, gravitational-wave, and (potential) dark-matter observations~\cite{LIGOScientific:2016aoc,KAGRA:2019htd,IceCube:2017amx}. Due to the finite mass of dark matter, BDM signals from nearby supernovae would be detected with a characteristic delay following the arrival of the supernova neutrino burst. Moreover, the correlation between neutrino and BDM signals can serve as a valuable probe for identifying dark-matter properties, including its mass scale.

To convey our ideas efficiently, this paper is organized as follows. In Section~\ref{sec:SNeBDM}, we define our benchmark dark matter scenario and review the mechanisms by which dark matter in this scenario can attain large boost factors from supernova explosions. Section~\ref{sec:detection} discusses the detection of supernova-induced BDM at large-volume neutrino detectors, detailing the analysis strategy along with key parameters and features of our benchmark neutrino experiments. Our main results, including the detection prospects of supernova-induced BDM signals at these detectors and their implications for multi-messenger astrophysics, are presented in Section~\ref{sec:results}. Finally, we provide our conclusions in Section~\ref{sec:conclusions}. 

\section{Boosted Dark Matter from Supernovae}\label{sec:SNeBDM}

To investigate the detection prospects of supernova-induced BDM at large-volume neutrino experiments, we consider a specific theoretical framework based on vector-portal dark matter models. For illustration, we assume fermionic dark matter, denoted by $\chi$ with mass $m_\chi$, interacting with Standard Model electrons via the exchange of a dark-sector gauge boson. We adopt the commonly studied case of a dark photon, $V_\mu$, which kinetically mixes with the Standard Model photon. The relevant interaction Lagrangian is given by \cite{Dutta:2020vop,Dutta:2024kuj}
\begin{eqnarray}
    -\mathcal{L}_{\rm int} \supset \epsilon e \bar{\chi} \gamma^\mu \chi V_\mu+g_D \bar{\chi}\gamma^\mu \chi V_\mu, \label{eq:lagrangian}
\end{eqnarray}
where $\epsilon$ is the kinetic mixing parameter, $e$ is the standard electromagnetic coupling, and $g_D$ parameterizes the coupling strength between $\chi$ and the dark photon.  

\subsection{Average diffuse BDM flux from core-collapse supernova events} \label{sec:DFLUX}

The fermionic dark matter produced in supernovae is emitted with a range of semi-relativistic velocities. As a result, a continuous flux of high-energy dark matter or supernova-induced BDM, arising from the overlapping emissions of different supernovae, is expected to reach Earth at any given time. To calculate the diffuse flux of supernova-induced BDM, we follow the approach outlined in Ref.~\cite{DeRocco:2019jti}, adopting the spatial supernova distribution model introduced in Ref.~\cite{adams2013observing}, which characterizes the rate of core-collapse supernovae (i.e., $dN_{\rm SN}/dt$) through a double-exponential function:
\begin{eqnarray}
\frac{dN_{\text{SN}}}{dt} = A \, e^{-r/R_d} \, e^{-|z|/H}, \label{eq:SNeDist}
\end{eqnarray}
where $A$ is a normalization parameter, $r$ represents the radial distance from the galactic center, and $z$ denotes the vertical offset from the midplane of the galaxy. For Type II supernovae, the two scale parameters, $R_d$ and $H$ are set to be 2.9~kpc and 95~pc, respectively, as specified in Ref.~\cite{adams2013observing}. Assuming an estimated galactic supernova rate of one event every 50 years, the normalization constant is determined to be $A = 0.00208 \, \text{kpc}^{-3} \, \text{yr}^{-1}$. Earth's radial distance from the galactic center $R_E$ and vertical offset $z_E$ are taken to be 8.7~kpc and 24~pc, respectively. 

The BDM flux contribution from a single supernova event diminishes with the square of its distance from Earth. To calculate the total diffuse BDM flux, we integrate over the distribution of supernovae in Eq.~\eqref{eq:SNeDist}, weighting each contribution by $1/(\vec{r} - \vec{R}_E)^2$, where $\vec{r}$ and $\vec{R}_E$ denote the three-dimensional positions of a given supernova and Earth in galactic coordinates, respectively. The resulting total flux $\Phi_\chi$ is
\begin{eqnarray}
\Phi_\chi = N_\chi \int_0^{z_{\text{max}}} dz \int_0^{2\pi} d\theta \int_0^{R_{\text{max}}} dr \frac{dN_{\text{SN}}}{dt} \frac{1}{4\pi(\vec{r} - \vec{R}_E)^2},
\label{eq:DINT}
\end{eqnarray}
where the quantity $N_\chi$ represents the total number of BDM particles produced by a single supernova, specifically those escaping the protoneutron star. 
We adopt the values calculated in Ref.~\cite{DeRocco:2019jti} based on Monte Carlo simulations. The value of $N_\chi$ depends on the dark-matter mass $m_\chi$ and its relevant coupling, which is characterized by an effective coupling parameter $y$, defined as 
\begin{eqnarray}
    y = \epsilon^2 \alpha_D \left( \frac{m_\chi}{m_V} \right)^4\,,
\end{eqnarray}
where $\alpha_D=g_D^2/(4\pi)$. Evaluating the integral in Eq.~\eqref{eq:DINT} at Earth's location with $R_{\max}=R_E$ and $z_{\max}=z_E$ where $R_E=8.7$~kpc and $z_E=24$~pc gives the expected BDM flux at Earth from the galactic supernova population:\footnote{The number reported in Ref.~\cite{DeRocco:2019jti} has typos. We have confirmed this with the authors.}
\begin{eqnarray}
\Phi_\chi =N_\chi(m_\chi,y) \times 2.11 \times 10^{-55} \, \text{cm}^{-2} \, \text{s}^{-1},
\end{eqnarray}
 where the normalization factor depends on the BDM mass $m_{\chi}$ and the coupling $y$. Finally, the BDM energy spectrum can be well approximated by a Fermi-Dirac distribution at the temperature $T$ corresponding to its thermal decoupling from the protoneutron star, leading to the following differential flux at the production point:
\begin{equation}
    \left. \frac{d\Phi_\chi}{d E}\right|_{\rm SN} = N_\chi(m_\chi,y)\left( \frac{E^2-m_\chi^2}{\text{exp}(E/T)+1} \right) \left( \int_{m_\chi}^\infty \frac{E^2-m_\chi^2}{\text{exp}(E/T)+1} \right)^{-1}.
\end{equation}
Assuming that the produced BDM particles do not significantly interact with the galactic medium during their propagation to Earth, the differential BDM flux at Earth is given by 
\begin{equation}
   \left. \frac{d\Phi_\chi}{d E} \right|_E = \Phi_\chi \left( \frac{E^2-m_\chi^2}{\text{exp}(E/T)+1} \right) \left( \int_{m_\chi}^\infty \frac{E^2-m_\chi^2}{\text{exp}(E/T)+1} \right)^{-1}.
\end{equation}
In our analysis, we set the temperature to be $T=30$~MeV.

\begin{figure}[ht!]
    \centering
    \begin{subfigure}[b]{0.65\textwidth}
        \centering
        \includegraphics[width=\textwidth]{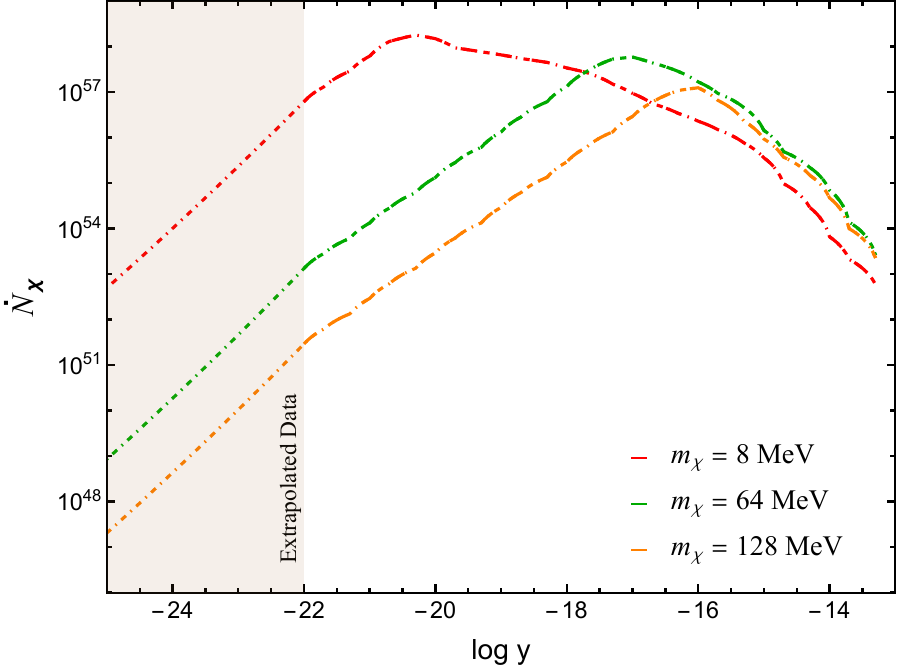}
    \end{subfigure}
    \captionsetup{width=1\linewidth, justification=justified, skip=2pt}
    \caption{Production rate of fermionic BDM originating from a core-collapse supernova event, considering different dark matter masses. The analysis extends the explored range of the coupling parameter to values smaller than those presented in Ref.~\cite{DeRocco:2019jti} (light-brown-shaded region). To achieve this, we extrapolate the flux by assuming a constant slope on a log-log scale, based on the trend observed in the data from Ref.~\cite{DeRocco:2019jti}, beyond their reported coupling limits. } 
    \label{fig:flux} 
\end{figure}

It is instructive to examine the strength of BDM production from a single core-collapse supernova to build intuition about its detection prospects. 
In Figure~\ref{fig:flux}, we show the instantaneous BDM particles $\dot{N}_\chi(=\partial N_\chi/\partial t)$ at a supernova as a function of the coupling parameter $y$ for three example $m_\chi$ values. The flux data is taken from Ref.~\cite{DeRocco:2019jti}, and we have extended the flux values to smaller coupling parameters, assuming an approximately constant slope in this range. We then evaluate $N_\chi(m_\chi,y)$ as $N_\chi =\dot{N}_\chi \Delta t$, corresponding to the total number of BDM particles emitted per supernova over a duration $\Delta t = \log(10) \, \text{seconds}$.

\subsection{BDM flux from a single core-collapse supernova event}

The BDM flux from a single core-collapse supernova event, i.e., a point-like source, can be estimated by the formalism discussed in the previous section. First of all, the spatial supernova distribution is simply given by a Dirac delta function, 
\begin{equation}
    \frac{dN_{\rm SN, src}}{dt}=\delta(\vec{r}-\vec{r}_{\rm src}),
\end{equation}
where $\vec{r}_{\rm src}$ is the location of the source supernova of interest. Substituting this expression into Eq.~\eqref{eq:DINT} simplifies the integration, resulting in the total BDM flux at Earth,
\begin{equation}
    \Phi_{\chi,{\rm src}}=\frac{N_\chi(m_\chi,y)}{4\pi(\vec{r}_{\rm src}-\vec{R}_E)^2},
\end{equation}
and the differential BDM flux at Earth,
\begin{equation}
    \left. \frac{d\Phi_{\chi,{\rm src}}}{dE} \right|_E =\frac{N_\chi(m_\chi,y)}{4\pi(\vec{r}_{\rm src}-\vec{R}_E)^2} \left( \frac{E^2-m_\chi^2}{\text{exp}(E/T)+1} \right) \left( \int_{m_\chi}^\infty \frac{E^2-m_\chi^2}{\text{exp}(E/T)+1} \right)^{-1}. \label{eq:fluxpointsource}
\end{equation}
Note that, not surprisingly, the flux depends only on the distance between Earth and the source supernova of interest. Since we consider Betelgeuse as an illustrative example for this scenario, as mentioned earlier, we have $(\vec{r}_{\rm src}-\vec{R}_E)^2=(197~{\rm pc})^2$. 

\section{Detection of Boosted Dark Matter at Neutrino Detectors}\label{sec:detection}

We are now in the position to discuss how the supernova-induced BDM can be detected in our benchmark large-volume neutrino detectors. We begin by outlining the key features of the detectors that are relevant for estimating signal sensitivity. This is followed by a discussion of the direct detection mechanism and the expected rate of signal events, based on the model described in the previous section.

\subsection{Benchmark neutrino detectors}

As discussed in the previous section, BDM produced by supernova bursts can be as energetic as the sub-GeV scale, imparting tens of MeV to GeV-scale recoil energies to target particles in a detector. Many large-volume neutrino detectors operate within this ideal energy range to register BDM-induced scattering events. Additionally, their extensive operational lifetimes and high-duty cycles maximize exposure, providing an advantageous route for rare event detection. 

Various large-volume neutrino experiments are currently operational or planned for the near future. In this study, we select three benchmark experiments with target volumes ranging from sub-kiloton to sub-megaton scales: DUNE, HK, and JUNO. For reference in subsequent sections, we outline some of the key detector specifications for these benchmark experiments.  

\begin{enumerate}
    \item \textbf{DUNE}~\cite{DUNE:2018hrq, DUNE:2018mlo, DUNE:2020lwj, DUNE:2020ypp, DUNE:2020txw}: The far detector of the Deep Underground Neutrino Experiment (DUNE) was originally planned to comprise four Liquid Argon Time Projection Chamber (LArTPC) modules and will be situated approximately 1,500 meters underground ($\approx 4,300$ meters water equivalent) at the Sanford Underground Research Facility in South Dakota, USA. At this depth, the expected cosmic-ray rate is $\sim 0.6~{\rm m}^{-2}{\rm hr}^{-1}$. The first two far-detector modules, both LArTPC-based, are expected to be operational by late 2029, while the remaining two modules, including one LArTPC-based and another designated as a module of opportunity~\cite{DUNE:2024wvj}, are scheduled to begin scientific operations in early 2031. Each module will contain a total liquid argon (LAr) mass of about 17.5 kilotons (or 17.5 kt LAr-equivalent), with a fiducial mass of at least 10 kilotons (or 10 kt LAr-equivalent). These detectors are designed to provide excellent angular resolution, strong particle identification capabilities, and a relatively low energy threshold. For this study, we assume that all four modules are LArTPC-based, as any variation does not significantly affect our conclusions, provided the total detector mass remains at $4\times 17.5$~kt LAr-equivalent.

    \item \textbf{Hyper-Kamiokande}~\cite{Abe:2011ts,Hyper-Kamiokande:2016srs,Hyper-Kamiokande:2018ofw}: HK is a next-generation underground water Cherenkov experiment, building on the highly successful operation of Super-Kamiokande.  The T2HK/T2HKK long-baseline neutrino experiment will use HK as its far detector for the upgraded J-PARC beam. The experiment will consist of two detectors, each featuring a cylindrical water tank, surrounded by about 40,000 photomultiplier tubes (PMTs). Each tank has 60 meters tall (with a fiducial height of 51.8 meters) and 74 meters in diameter (67.8 meters fiducial). They hold a total of 258 kilotons of water of which 187 kilotons are the fiducial volume. The first detector will be located at the Tochibora mine, near the current Super-Kamiokande site in Japan, at a depth of 650 meters $\approx 1,750$ meters water equivalent), where the expected cosmic-ray rate is $\sim 27~{\rm m}^{-2}{\rm hr}^{-1}$. The second detector is currently under consideration for construction beneath Mt. Bisul or Mt. Bohyun in Korea, with an overburden of about 1,000 meters ($\approx 2,700$ meters water equivalent), resulting in a reduced cosmic-ray rate of $\sim 5.7~{\rm m}^{-2}{\rm hr}^{-1}$. The first detector in Japan is expected to begin operations in 2027. For electron detection, HK will achieve a relatively low energy threshold of $\sim 5$~MeV, although a significantly higher energy threshold is required to attain good angular resolution. For this study, we consider the first detector only.  

 \item    \textbf{JUNO}~\cite{JUNO:2015zny,JUNO:2015sjr,Guo:2019vuk}: The Jiangmen Underground Neutrino Observatory (JUNO) in Kaiping, Jiangmen, China is an advanced neutrino experiment utilizing liquid scintillator and is expected to begin operations in 2025.  The JUNO central detector consists of a 35.4-meter-diameter acrylic sphere filled with 20 kilotons of linear alkylbenzene-based liquid scintillator, making it the largest liquid scintillator detector ever constructed. This sphere is surrounded by approximately 17,600 20-inch PMTs and 25,600 3-inch PMTs, providing an unprecedented light collection efficiency, an energy threshold as low as 0.1~MeV, and an excellent \%-level energy resolution. The detector is located 700 meters underground (approximately 1,800 meters water equivalent) to reduce cosmic-ray backgrounds. JUNO's primary physics goals include precision measurements of neutrino oscillation parameters, supernova neutrino detection, and searches for new physics. The experiment is expected to achieve a low energy threshold, making it highly sensitive to a wide range of astrophysical and exotic neutrino sources. For this study, we assume JUNO's full fiducial mass of 20 kilotons and its nominal detector performance, as any small variations in operational parameters are not expected to significantly impact our conclusions.
\end{enumerate}

As briefly discussed earlier, while direct detection experiments are sensitive to nuclear recoils, the above benchmark neutrino experiments can search for supernova-induced BDM signals through electron scattering, benefiting from their large target volumes and low thresholds for electron recoils. This provides complementary information in the broader search for supernova-induced BDM. The number of target electrons (henceforth denoted by $N_T$) varies across neutrino experiments based on their distinct designs. The massive HK detector has the largest count, with approximately $6.27 \times 10^{34}$ electrons. DUNE follows with a significant, yet smaller, total of about $1.084 \times 10^{34}$ electrons. JUNO has a comparatively lower count of roughly $6.95 \times 10^{33}$ electrons, reflecting how the target scale is optimized for specific scientific objectives. DUNE’s minimum detectable energy threshold for supernova-induced neutrinos is expected to be around 5 MeV, set by practical considerations of spallation backgrounds~\cite{DUNE:2020zfm, Zhu:2018rwc}. HK adopts a similar threshold of approximately 5 MeV for supernova-induced neutrino detection to suppress backgrounds~\cite{Hyper-Kamiokande:2021frf}, although lower thresholds ($3.5-4.5$ MeV) are considered for solar neutrino measurements. JUNO, optimized for reactor antineutrinos, operates with a threshold near 0.1 MeV~\cite{Fang:2019lej, JUNO:2021vlw}, and aims for a higher threshold of about 2 MeV for solar neutrinos, depending on the level of radioactive contaminants. These numbers are summarized in Table~\ref{tab:threshold}. Based on these considerations, we adopt a threshold of $5$ MeV for DUNE and HK, and $0.1$ MeV for JUNO to assess their maximum detection potentials. More conservative thresholds are also explored later in our analysis.

\begin{table}[t]
\centering
\small 
\begin{tabular}{|c|c|c|>{\centering\arraybackslash}m{4.2cm}|}
\hline
\textbf{Experiment} & \textbf{Material} & \boldmath$N_T$ \textbf{(Electrons)} & \textbf{Energy Thresholds} \\
\hline
HK & Water (H$_2$O) & $6.27 \times 10^{34}$ & $\sim5\,\mathrm{MeV}$~(supernova)~\cite{Hyper-Kamiokande:2021frf}; \newline $3.5$--$4.5\,\mathrm{MeV}$ (solar)\\ 
\hline
DUNE & Liquid Argon (Ar) & $1.084 \times 10^{34}$ & $\sim5\,\mathrm{MeV}$~(supernova) ~\cite{DUNE:2020zfm,Zhu:2018rwc} \\
\hline
JUNO & Liquid Scintillator (C$_9$H$_{12}$) & $6.95 \times 10^{33}$ & $\sim0.1\,\mathrm{MeV}$ (reactor); \newline $\sim2\,\mathrm{MeV}$ (solar)
~\cite{Fang:2019lej,JUNO:2021vlw} \\ 
\hline
\end{tabular}
\caption{The number of target electrons $N_T$ and energy thresholds for our benchmark neutrino experiments. Thresholds vary depending on the neutrino sources and background suppression capabilities. }
\label{tab:threshold}
\end{table}

\subsection{Scattering between BDM and electron} 
\label{sec:DME}

Given the interaction Lagrangian in Eq.~\eqref{eq:lagrangian}, an incoming BDM particle can manifest itself as an electron recoil via a $t$-channel exchange of the dark photon $V$. The matrix element describing the scattering process $\chi e^- \to \chi e^-$ is written as
\begin{equation}
    i \mathcal{M} = \frac{i \epsilon e g_{D}}{t - m_V^2}\bar{u}(p')\gamma^\mu u(p)\bar{u}(k')\gamma_\mu u(k),
\end{equation}
where $t$ is the Mandelstam variable defined as $t=(p-p')^2$.
The initial and final momenta of BDM particles and electrons in the laboratory frame are given by

\begin{equation}
    p^\mu = \begin{bmatrix} E_\chi \\ \vec{p} \end{bmatrix}, \hspace{0.5cm} k^\mu = \begin{bmatrix} m_e \\ \vec{0} \end{bmatrix}, \hspace{0.5cm} p'^\mu = \begin{bmatrix} E'_\chi \\ \vec{p'} \end{bmatrix}, \hspace{0.5cm} k'^\mu = \begin{bmatrix} E_{r,e} \\ \vec{k}' \end{bmatrix},
\end{equation}
where $E_\chi$ represents the energy of the incoming dark-matter particle, while $E_{r,e}$ denotes the energy of the recoiled electron. The squared amplitude of BDM and electrons can be written in spin-averaged form
\begin{equation}
    \overline{|\mathcal{M}|^2} = \frac{8\,\epsilon^2e^2g_{D}^2m_e}{(2m_e\{m_e-E_{r,e}\} - m_V^2)^2} \left[m_e\{E_\chi^2+(m_e+E_\chi-E_{r,e})^2 \} + (m_e^2 + m_\chi^2)(m_e-E_{r,e})\right].
\end{equation}
The differential cross section in the laboratory frame can be expressed as
\cite{Dutta:2020vop,Dutta:2024kuj}
\begin{align}
    \frac{d\sigma}{dE_{r,e}} &= \frac{m_e}{8\pi \lambda(s,m_e^2,m_\chi^2)}\overline{|\mathcal{M}|^2} \\
    &= \frac{\epsilon^2 e^2 g_{D}^2 m_e^2}
    {\pi \lambda(s,m_e^2,m_\chi^2)(2m_e[m_e-E_{r,e}] - m_V^2)^2} \nonumber \, \times \\
    &\hspace{2.5 cm} \left[ m_e \left\{ E_\chi^2 + (m_e + E_\chi - E_{r,e})^2 \right\} 
    + (m_{e}^2 + m_{\chi}^2)(m_{e} - E_{r,e}) \right],
\end{align}
where $\lambda(x,y,z) = (x-y-z)^2-4yz$ and $s=m_e^2+2E_\chi m_e +m_\chi^2$. The kinematically allowed values of the electron recoil energy are determined by energy-momentum conservation. The maximum (minimum) allowed recoil energy $E_e^+$ ($E_e^-$) is \cite{Kim:2016zjx}
\begin{equation}
E_e^\pm = \frac{(s + m_{e}^2 - m_\chi^2)(E_{\chi}+m_e) \pm \sqrt{\lambda(s, m_{e}^2, m_\chi^2)}\,p_\chi}{2s} \,,
\label{eq:maxrecoil}
\end{equation}
where $p_\chi$ is the momentum of the BDM particle, given by $p_{\chi} = \sqrt{E_{\chi}^2 - m_{\chi}^2}$.

In summary, the outlined framework offers a consistent and comprehensive description of elastic BDM scattering off electrons. The underlying interaction is governed by a Lagrangian in which the dark-matter particle couples to a dark-sector mediator that kinetically mixes with the Standard Model photon. The scattering kinematics are fully specified by the Mandelstam variable $s$, the K\"{a}ll\'{e}n function $\lambda$, and the explicit expression for the additional kinematic factor $\mathcal{K}(E_\chi, m_e, m_\chi, E_r)$, which encapsulates the energy and mass dependence of the process.  Importantly, the formulation retains the full momentum-transfer dependence of the mediator propagator, ensuring precise modeling of the recoil spectra. This level of detail is essential for making reliable predictions in the context of BDM detection at electron-sensitive experiments.

\subsection{Background considerations}
Since our benchmark detectors are, or will be, located deep underground, one expects that the background contamination from cosmic rays (primarily from cosmic muons) would be insignificant. Nevertheless, the annual flux is not negligible, necessitating a more careful estimate.

The total muon fluxes expected at DUNE, HK, and JUNO detector locations are approximately $4\times 10^{-5}~{\rm m}^{-2}{\rm sr}^{-1}{\rm s}^{-1}$, $4\times 10^{-4}~{\rm m}^{-2}{\rm sr}^{-1}{\rm s}^{-1}$, and $2\times 10^{-3}~{\rm m}^{-2}{\rm sr}^{-1}{\rm s}^{-1}$, respectively~\cite{ParticleDataGroup:2024cfk}. Based on these flux values, we estimate that about $10^7$ muons, $4.7\times 10^8$ muons, and $2.5\times 10^8$ muons pass annually through the DUNE-40kt, HK-187kt, and JUNO-20kt detector volumes, respectively.  The outer detector of HK is used to detect the background cosmic muons and prevent them from affecting the data. The tagging efficiency---and therefore the muon rejection efficiency---is expected to exceed 99.9\%~\cite{Abe:2011ts,HKweb}. JUNO's central detector is immersed in a water pool equipped with a Cherenkov detection system, which is expected to identify cosmic muons with an efficiency of 99.8\%~\cite{JUNO:2021vlw}. In addition, the top tracker module provides further cosmic muon rejection capability~\cite{JUNO:2021vlw}. In contrast, the DUNE far detectors do not include dedicated muon tagger modules. However, the photon detection system, combined with the excellent tracking capabilities of the LArTPC, enables muon tagging with an efficiency approaching 100\%. In order for a muon to mimic a signal event, it must evade detection at the muon tagger modules or the active volume, sneak into the detector fiducial volume, and radiatively emit a photon that subsequently converts into an $e^-e^+$ pair~\cite{Chatterjee:2018mej,DeRoeck:2020ntj}. However, such emission becomes significant only above 100 GeV~\cite{Workman:2022ynf}, where the related muon flux is highly suppressed underground and the resulting electron energy deposits are much larger than those typically expected from supernova-induced BDM. Based on all these considerations, the cosmic-muon-initiated backgrounds are expected to be insignificant or even negligible. 

The primary irreducible background arises from atmospheric neutrinos. Quasi-elastic scattering of electron-flavor neutrinos, where the outgoing nucleons escape detection, can potentially mimic the signal events. The neutral-current contributions are typically subdominant, generally measured to be about $10-50\%$ of the corresponding charged-current contributions, depending on the interaction channel, neutrino energy, and target material~\cite{Formaggio:2012cpf}. Therefore, we will focus on the $\nu_e/\bar{\nu}_e$ charged-current events for illustrative background estimates.

\begin{figure}[t]
    \centering
    \includegraphics[width=0.55\textwidth]{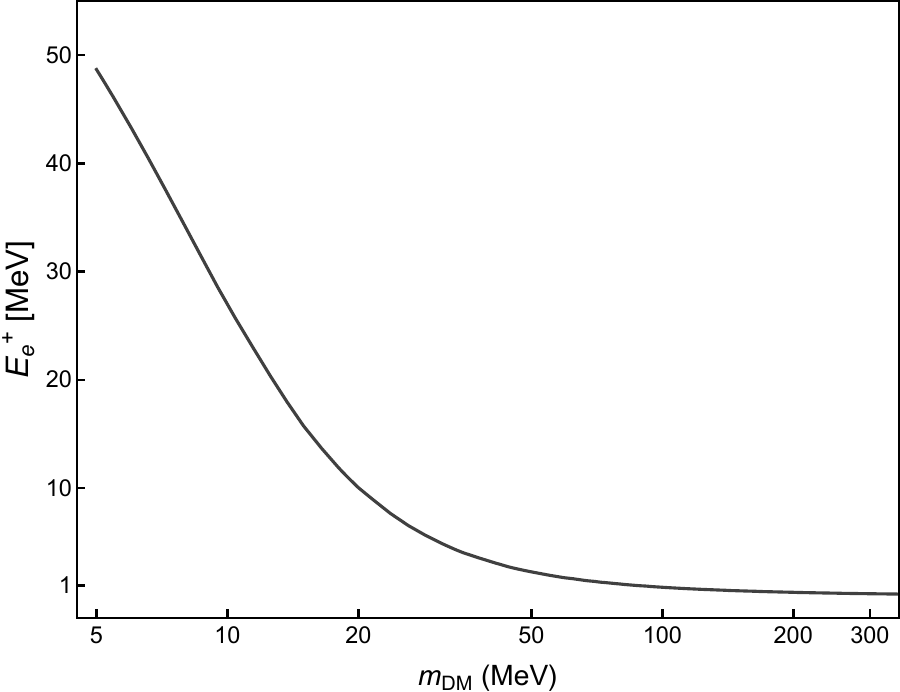}    
    \caption{Maximum electron recoil energy corresponding to the most probable BDM energy for each mass, shown as a function of the BDM mass.} 
    \label{fig:maxenergy}
\end{figure}

To estimate the event rate, we combine the differential atmospheric $\nu_e/\bar{\nu}_e$ flux at the SK site, as calculated in Ref.~\cite{Honda:2015fha}, with the neutrino scattering cross sections from Ref.~\cite{Formaggio:2012cpf}. Using these inputs, the number of events for the relevant channels can be calculated as a function of the incident neutrino energy, as reported in Ref.~\cite{DeRoeck:2020ntj}. Since our benchmark detector sites are located at similar latitudes and depths as the SK site, this serves as a reasonable approximation.
Based on the data in Ref.~\cite{DeRoeck:2020ntj}, we estimate that approximately 37 events are expected per kt$\cdot$yr, ignoring energy thresholds. However, typical supernova-induced BDM is not energetic enough to deposit more than $50-100$~MeV, depending on the mass of BDM. Figure~\ref{fig:maxenergy} shows the kinematically allowed maximum recoil energy as a function of BDM mass. To determine this value, we identify the most probable BDM energy for each mass and substitute it into Eq.~\eqref{eq:maxrecoil}. The maximum recoil energy remains below 50~MeV across the BDM mass range, as expected just before. Notably, it decreases with increasing BDM mass due to the relatively low decoupling temperature of around 30~MeV in the proto-neutron star, which limits the boost of heavier BDM particles. Additionally, heavier BDM species are produced less abundantly, further reducing detection prospects. 
These observations suggest that an upper energy cut will further reduce the background events induced by atmospheric neutrinos.  For example, less than $5\%$ of quasi-elastic scattering events are produced from $\nu_e/\bar{\nu}_e$  neutrinos that have energies below $100$~MeV.

When it comes to the BDM signals from a single supernova burst, signal isolation can be enhanced by utilizing timing and directionality information to further distinguish signal events from potential backgrounds. As we will discuss in Section~\ref{sec:multimessgenger}, following a nearby supernova explosion, neutrinos and photons arrive at the detector first,\footnote{Though photons reach Earth slightly later due to their electromagnetic interaction nature.} with BDM signals arriving later due to the finite mass of the BDM particles. Since photon and neutrino signals can be used to locate the supernova, the source direction of the BDM can be reconstructed by tracking recoil electrons with high angular resolution. For example, DUNE achieves an angular resolution of $\sim 1^\circ$~\cite{DUNE:2018hrq,DUNE:2018mlo,DUNE:2020lwj,DUNE:2020ypp,DUNE:2020txw}, while SK achieves between $25^\circ$ and $3^\circ$ for electrons in the $10-100$ MeV range~\cite{Fukuda:1998mi,Super-Kamiokande:2010tar,Super-Kamiokande:2014ndf,Super-Kamiokande:2015qek}. HK is expected to improve upon SK's performance with enhanced angular resolution.  On the other hand, the JUNO detector uses a liquid scintillator that has a limited capacity to trace the path of recoil electrons.However, a recent study based on various machine learning techniques has demonstrated the potential of reconstructing electron recoil tracks with an angular resolution of $\sim20^\circ- 30^\circ$~\cite{Yang:2023rbg}, although this performance was shown for GeV-scale $\nu_e$-induced electron recoils. 

All the considerations above clearly indicate that more precise background estimates will require dedicated studies that incorporate detailed detector response modeling, which lies beyond the scope of this work. Additionally, subdominant backgrounds not discussed above may become considerable or even significant in certain parameter regions, while evolving detector capabilities may help address these challenges. Rather than proposing detailed signal detection strategies optimized for various scenarios and experimental environments, in our sensitivity study, we present contours corresponding to 2.3 [i.e., the number of signal events corresponding to a 90\% confidence level (C.L.), assuming negligible background], 10, 100, and 1,000 signal events. We anticipate that these illustrate the detection potential of supernova-induced BDM signals at varying strengths, while leaving detailed detector response studies---particularly those related to directionality---to the experimental collaborations.

\section{Results}\label{sec:results}

In this section, we discuss the expected event rates of supernova-induced BDM at our benchmark neutrino detectors. 
For the dark-matter model under consideration, a BDM particle reaching a detector can interact with either electrons or nuclei/nucleons in the detector material. If BDM interacts with the nucleus, it can manifest itself as a nuclear recoil, which existing WIMP detectors are well-equipped to observe~\cite{DeRocco:2019jti}. However, as discussed in the introductory section, in large-volume neutrino detectors, typical supernova-induced BDM is not energetic enough to surpass the threshold for nucleon recoils but can produce observable electron recoils above the MeV scale. For a neutrino detector, the total expected number of BDM events during the observation time $T_{\rm obs}$ is given by
\begin{equation}
    N_{\rm sig} = N_e \, T_{\rm obs} \int_{E_{\rm min}}^{E_{\rm max}}dE_{\rm r,e} \int_{m_\chi}^{\infty}dE \frac{d\sigma}{dE_{\rm r,e}}\frac{d\Phi_\chi}{dE}\,,  
\end{equation}
where $N_e$ is the total number of target electrons within the detector fiducial volume, $d\sigma/dE_{\rm r,e}$ is the differential BDM-electron cross section described in Section~\ref{sec:DME}, and $d\Phi_\chi/dE$ is the differential flux of BDM particles discussed in Section~\ref{sec:SNeBDM}. The relevant electron recoil energy range is determined by $E_{\min}$ and $E_{\max}$:
\begin{eqnarray}
    E_{\min}&=&\max[E_{\rm th},~E_e^-], \\
    E_{\max}&=& E_e^+,
\end{eqnarray}
where $E_e^\pm$ is defined in Eq.~\eqref{eq:maxrecoil}, and $E_{\rm th}$ denote the detector’s energy threshold.

\begin{figure}[t]
    \centering
    \begin{subfigure}[b]{0.49\textwidth}
        \centering
        \includegraphics[width=\textwidth]{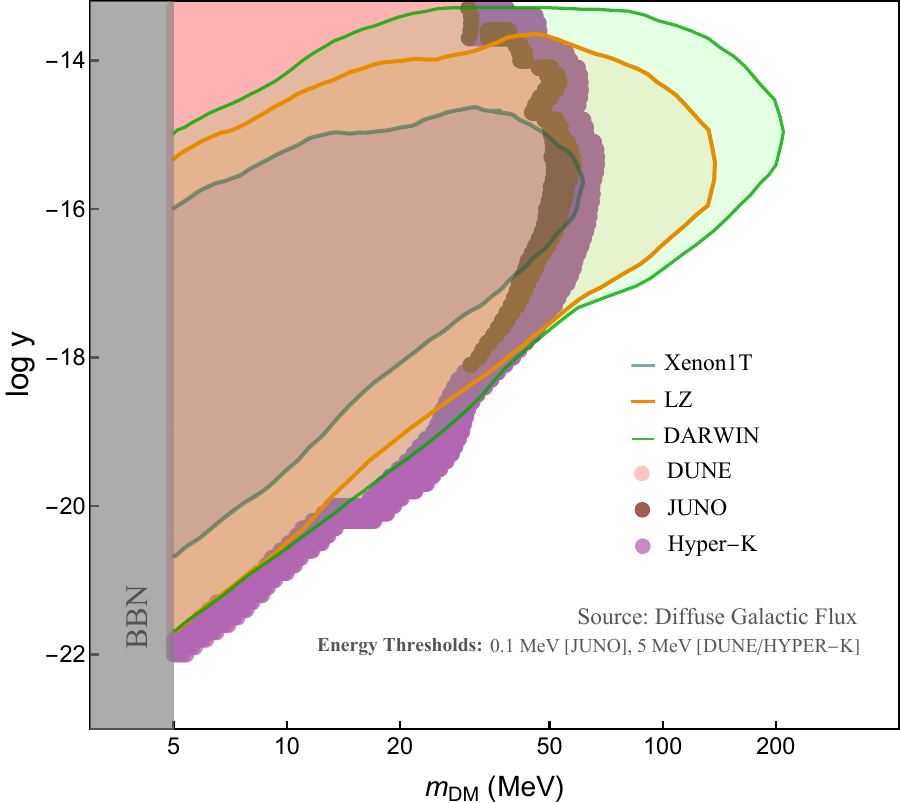}
    \end{subfigure}
    \begin{subfigure}[b]{0.49\textwidth}
        \centering
        \includegraphics[width=\textwidth]{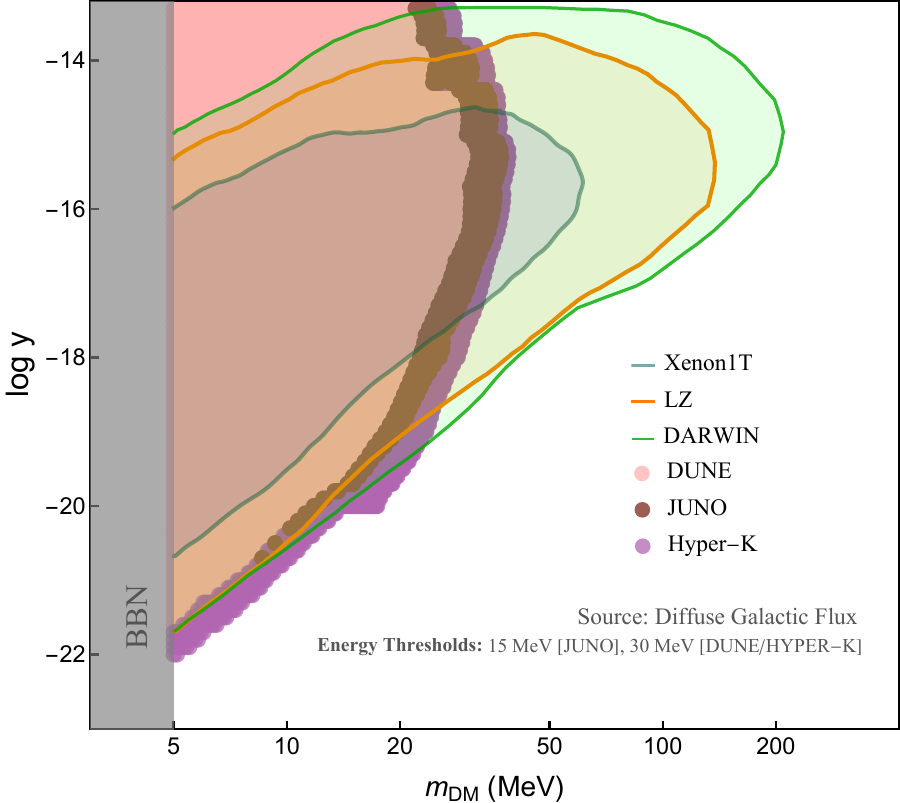}
    \end{subfigure}
    \captionsetup{width=1\linewidth, justification=justified, skip=2pt}
    \caption{The 90\% C.L. sensitivity regions for DUNE (in red), JUNO (in red and brown), and HK (in red, brown, and purple) for a 10-year observation period. Left: The detector threshold for electron recoil is set to 5 MeV for DUNE and HK \cite{DUNE:2020zfm,Zhu:2018rwc,Hyper-Kamiokande:2021frf}, while for JUNO, it is set at 0.1 MeV \cite{Fang:2019lej,JUNO:2021vlw} as mentioned in Table \ref{tab:threshold}. Right: The threshold for electron recoil is set to 30 MeV for DUNE and HK, and to 15 MeV for JUNO. The sensitivity regions of direct dark-matter detection experiments are shown in green, orange, and cyan, taken from Ref.~\cite{DeRocco:2019jti}, with a detector threshold for nuclear recoil set at 2.5 keV. The dark-matter mass below $5$ MeV is excluded by BBN~\cite{Krnjaic:2019dzc}.}
    \label{fig:sensitivity} 
    \vspace{0.5cm}
    \begin{subfigure}[b]{0.49\textwidth}
        \centering
        \includegraphics[width=\textwidth]{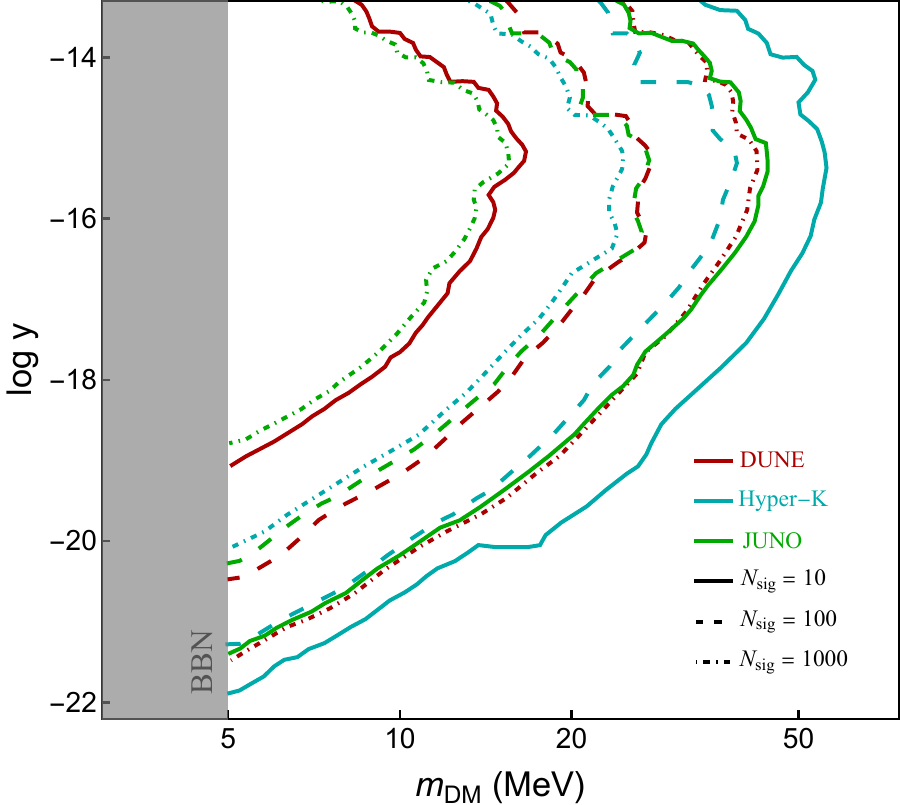}
    \end{subfigure}
    \begin{subfigure}[b]{0.49\textwidth}
        \centering
        \includegraphics[width=\textwidth]{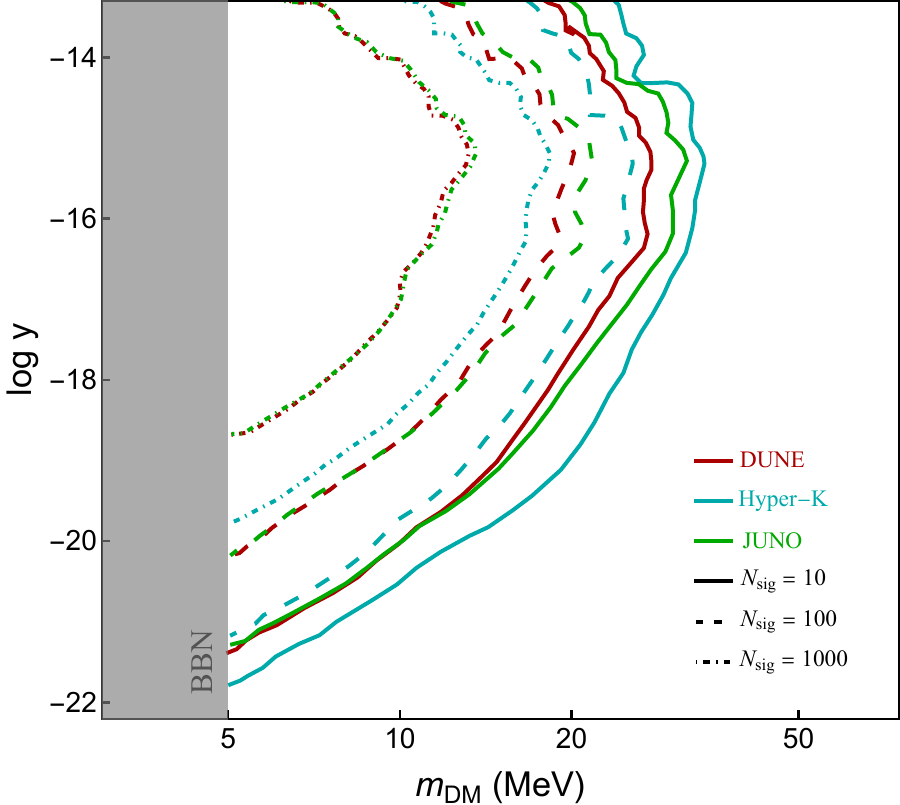}
    \end{subfigure}
    \captionsetup{width=1\linewidth, justification=justified, skip=3pt}
    \caption{Event counts for a diffused flux of supernova-induced BDM at DUNE, JUNO, and HK over a 10-year observation period. Left: the detector threshold for electron recoil is set to 5 MeV for DUNE and HK, while for JUNO, it is set at $0.1$ MeV. Right: The threshold for electron recoil is set to 30 MeV for DUNE and HK, and 15 MeV for JUNO. The dark-matter mass below $5$ MeV is excluded by BBN~\cite{Krnjaic:2019dzc}.}
    \label{fig:Nevents} 
\end{figure}

In the following subsection, we present our sensitivity estimates obtained for the diffuse galactic flux, originating from the combined emission of numerous galactic supernovae, and BDM signals from a nearby isolated supernova, using Betelgeuse as an example. The sensitivity regions for both scenarios are illustrated in Figures \ref{fig:sensitivity}, \ref{fig:Nevents}, and \ref{fig:betelB}, respectively.

\subsection{Diffuse galactic flux of supernova-induced BDM}
Dark matter particles produced in a supernova possess a semi-relativistic velocity with an $\mathcal{O}(1)$ spread, causing them to arrive at Earth significantly later than photons and neutrinos. However, this also means that BDM particles originating from different supernovae can arrive at Earth concurrently. The resulting flux was quantitatively analyzed in Section~\ref{sec:DFLUX}. The flux  depends on the dark matter mass $m_{\chi}$ and the effective coupling $y$ through the dark gauge boson. For large couplings, BDM becomes trapped within the proto-neutron star~\cite{DeRocco:2019jti}, causing the flux to be exponentially suppressed. Conversely, if the coupling is too small, dark-matter production itself is inefficient, also leading to a suppressed flux. Based on the results from Monte Carlo simulations in Ref.~\cite{DeRocco:2019jti}, we find that for each value of $ m_\chi $, there is an optimal value of the coupling parameter $ y_{\rm max}$ that maximizes the flux. 

To generate a detectable signal in a neutrino detector, the BDM must deposit enough energy to ensure that the electron recoil energy exceeds the detector's threshold, while remaining below the maximum energy probed by the instrument. Since most electron recoils from supernova-induced BDM are expected to lie near the detector’s threshold, the threshold energy $E_{\rm th}$ plays a key role in determining the event rate. 

Figure~\ref{fig:sensitivity} presents the estimated 90\% C.L. sensitivity to supernova-induced BDM signals at our benchmark neutrino detectors. Assuming negligible background levels, the energy thresholds could be lowered to 5 MeV for DUNE and HK, and to 0.1 MeV for JUNO, enabling us to estimate the maximum achievable sensitivity at each experiment. The results for this optimistic scenario are displayed in the left panel. In contrast, the right panel presents a more conservative sensitivity estimate, based on an energy threshold of 30 MeV for DUNE and HK, and 15 MeV for JUNO, again assuming a negligible background. In both cases, we assume a 10-year time exposure. For comparison purposes, we also show the sensitivity estimates of supernova-induced BDM at conventional WIMP detectors as reported in Ref.~\cite{DeRocco:2019jti}. In contrast to conventional WIMP detectors, neutrino detectors are more effective at probing lighter BDM masses. As the BDM mass increases, both the BDM-electron scattering cross section and the diffuse galactic flux decrease, resulting in a reduced event rate. Consequently, the sensitivity of neutrino detectors declines for heavier BDM.
Finally, as previously announced, we present in Figure~\ref{fig:Nevents} the contours corresponding to 10, 100, and 1,000 signal events for a 10-year exposure, reflecting different background assumptions. The left and right panels illustrate the estimates based on optimistic and conservative energy threshold choices, respectively.

\subsection{Multi-messenger signals from a single supernova event} \label{sec:multimessgenger}

 A  red supergiant located in the Orion constellation, Betelgeuse, is getting close to  the end of its life cycle and is located  nearly 642.5 light-years (or 197 parsecs) away from Earth~\cite{wheeler2023betelgeuse, saio2023evolutionary}. Its eventual supernova explosion offers a rare opportunity to detect BDM from a well-localized, point-like astrophysical source. In this context, both neutrinos and dark matter particles are expected to be emitted from a compact region, allowing the resulting BDM flux to be described as in Eq.~\eqref{eq:fluxpointsource}.

Supernova-produced BDM particles, due to the (semi-)relativistic velocities at which they are emitted, will arrive at Earth sometime after the photons and neutrinos. If the supernova is relatively close to Earth, the time difference between the arrival of the BDM and neutrinos can be quite small. Therefore, a single nearby supernova event offers an excellent opportunity for multi-messenger studies involving dark matter and neutrinos. Letting $\Delta T$ represent the time difference of arrival between the photons and the dark matter particles, we have
\begin{equation}
    \Delta T = \frac{D_S}{v_\chi} - \frac{D_S}{c},
    \label{eq:DeltaT}
\end{equation}
where $D_S$ is the distance of the source from Earth, $v_\chi$ is the velocity of the BDM particle, and $c$ is the speed of light.\footnote{Hereafter, we set the value of $c$ to 1.} The velocity of BDM particles depends on the mass parameter and the energy at which they are produced. For a dark-matter particle of mass $m_\chi$ and energy $E_\chi$, its velocity is given by
\begin{equation}
    {v_\chi^2} = 1-\frac{m_\chi^2}{E_\chi^2}.
    \label{eq:vME}
\end{equation}
At any given time, a dark matter particle of a specific mass arriving at the detector carries a fixed amount of energy. As a result, the number of observed events per unit time can be written as 

\begin{eqnarray}
    \frac{dN_{\rm sig}}{dt} = N_e\int_{-\infty}^{+\infty}dE \frac{d\Phi_\chi}{dE} \delta(E-E_\chi[t])\int_{m_\chi}^{\infty}dE_{\rm r,e}  \frac{d\sigma}{dE_{\rm r,e}}\,.
\end{eqnarray}

\begin{figure}[t]
        \centering
        \includegraphics[width=0.55\textwidth]{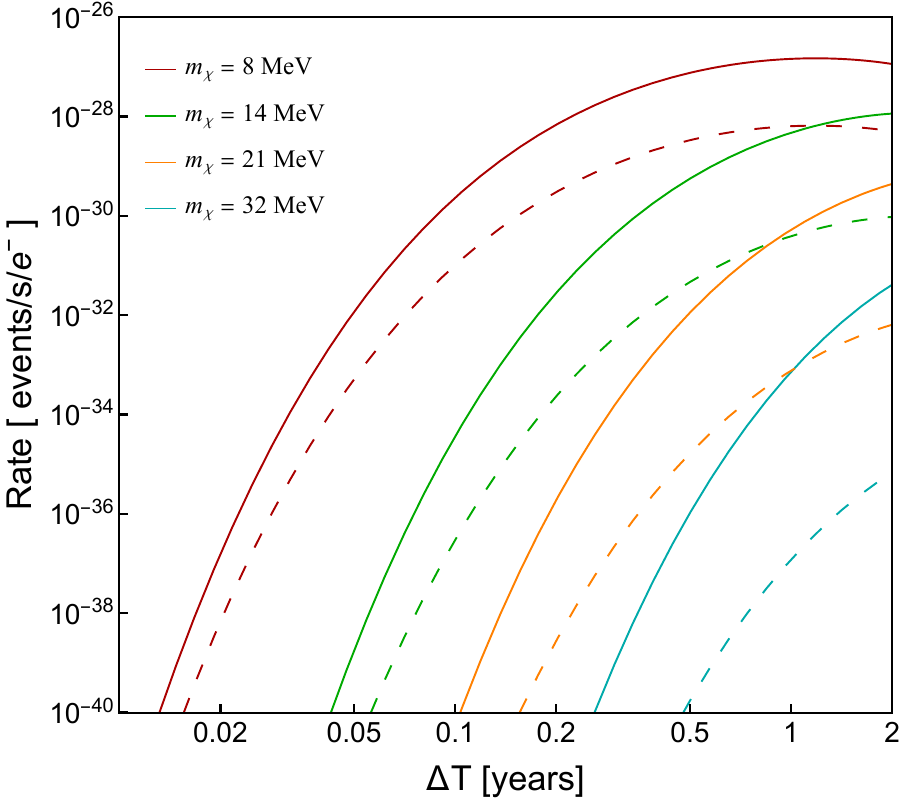}
    \captionsetup{width=1\linewidth, justification=justified, skip=2pt}
    \caption{The event rate of BDM particles from Betelgeuse per unit time per target electron. The solid lines correspond to $y=10^{-18}$, while the dashed lines correspond to $y=10^{-14}$.  The detector threshold is considered to be $30$~MeV in our estimation.}
    \label{fig:betelmultimessenger}
\end{figure}

In Figure~\ref{fig:betelmultimessenger}, we show the event rate of BDM at neutrino detectors as a function of time following neutrino detection, considering a range of masses and coupling strengths. Since the core temperature in a SN is approximately 30 MeV, dark matter particles with masses below this temperature are produced with significant boosts and arrive soon after the neutrinos. In contrast, heavier dark matter particles are less boosted and may take several years to come after the neutrinos. Therefore, unlike the diffuse galactic flux, which remains approximately constant over time, the dark matter flux from a nearby single source, such as Betelgeuse, can vary significantly. Hence, the total number of events expected for a neutrino detector during the observation period $T_{\text{obs}}$ is calculated by time-integrating the flux. 
\begin{eqnarray}
    N_{\rm sig} = N_e \int_0^{T_{\rm obs}}dt\int_{-\infty}^{+\infty}dE \frac{d\Phi_\chi}{dE} \delta(E-E_\chi[t])\int_{m_\chi}^{\infty}dE_{\rm r,e}  \frac{d\sigma}{dE_{\rm r,e}}\,.
\end{eqnarray}

\begin{figure}[t]
        \centering
        \includegraphics[width=0.49\textwidth]{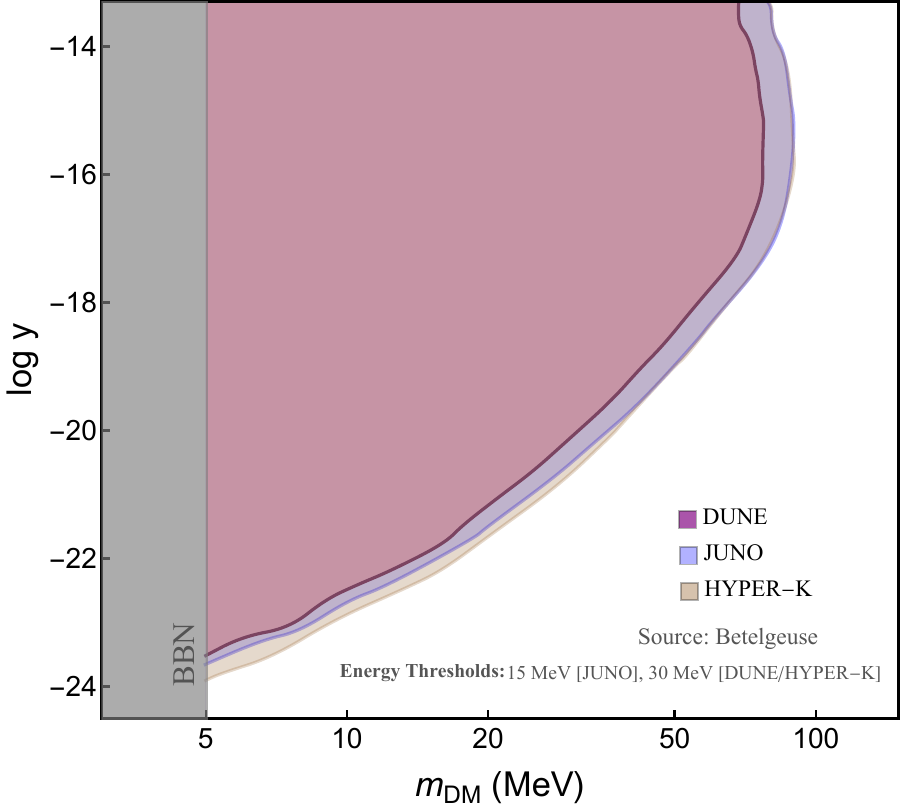}
    \includegraphics[width=0.5\textwidth]{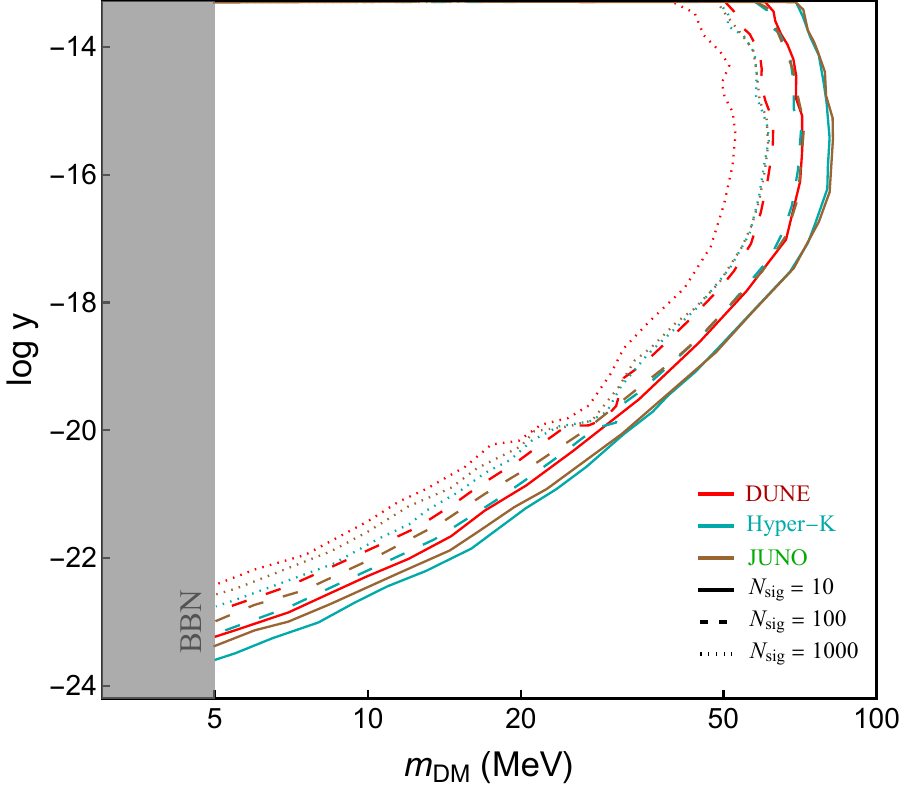}
    \captionsetup{width=1\linewidth, justification=justified, skip=2pt}
    \caption{Left: The sensitivity region for DUNE, JUNO, and HK. The flux is calculated based on the distance of Betelgeuse from Earth, which is $197$ parsecs. We consider an upper cut on the value of $y$ coupling to be smaller than $10^{-13}$ to avoid the trapping regime of dark matter by the supernova, see e.g., Ref. \cite{Chang:2018rso}. Right: Contours for different numbers of events.  The dark-matter mass below $5$ MeV is excluded by BBN~\cite{Krnjaic:2019dzc}.}
    \label{fig:betelB}
\end{figure}

The corresponding sensitivity at 90\% C.L. for a 10-year exposure, assuming negligible background, is shown in the left panel of Figure~\ref{fig:betelB}. As anticipated, the sensitivity decreases for relatively heavier dark matter particles, as they are less boosted and may require longer than the observation time to reach Earth. Finally, the right panel of Figure~\ref{fig:betelB} displays the contours corresponding to different background assumptions, i.e., 10, 100, and 1,000 signal events over a 10-year exposure. 

\section{Summary and Conclusions} \label{sec:conclusions}

We studied fermionic dark matter produced and boosted at supernovae that can be probed in next-generation neutrino detectors.
The BDM of this sort interacting via dark gauge bosons (e.g., dark photons), can be produced and accelerated in core-collapse supernovae. It can give a relatively large flux that might be observable at Earth. By considering the modeling of diffuse galactic supernovae and nearby sources like Betelgeuse, we predict event rates at DUNE, HK, and JUNO. Moreover, we investigated a multi-messenger approach that represents a sensitivity to unexplored dark-matter parameter space, especially in the MeV to sub-GeV mass range with feeble couplings.

Supernovae as natural sources for energetic particles can produce BDM efficiently via dark gauge boson interactions due to the extreme conditions. We have considered both diffuse flux as a background from past supernovae events and bursts from nearby events motivated by the simulation in Ref.~\cite{DeRocco:2019jti}. Producing energetic dark matter at supernovae can be observable in near-future sensitive dark matter and neutrino detectors. BDM from nearby supernova events could present excellent detection possibility due to larger flux and closer distance to Earth.

We have also looked into a possible multi-messenger approach for BDM emission and its detectability prospect at Earth. The time delay between neutrinos and BDM is related to the dark-matter mass and speed, providing a means to distinguish potential signals from either diffuse or point-like sources. Correlations with other messengers, such as gravitational waves, can further enhance the multi-messenger approach.

Furthermore, we have found that a wide range of dark-matter masses and interaction strengths coupled to the Standard Model can be explored through a dark gauge boson. We have shown our main sensitivity prospects in Figures~\ref{fig:sensitivity} and \ref{fig:Nevents} for diffuse sources and in Figure~\ref{fig:betelB} for point-like source at the location of Betelgeuse.  Using the characteristic features of current and future neutrino detectors, we have obtained projected sensitivity curves covering a range of dark-matter masses and effective couplings. A possible supernova event could either produce a BDM signal---potentially leading to the first non-gravitational detection of dark matter originating from the extreme environment of a supernova---or further improve existing constraints. Depending on the number of observed events, our sensitivity curves span dark-matter masses from $\sim 5$~MeV to $\sim 100$~MeV, with effective coupling values $y$ ranging from $\sim10^{-13}$ to $\sim 10^{-22}$, as shown in the figures. We have also compared our results with the sensitivity curves for dark-matter direct detection experiments presented in Ref.~\cite{DeRocco:2019jti}.

The high energy resolution and huge fiducial volume of DUNE, the continuous operation of JUNO, and the vast size of HK can improve complementary capabilities for probing BDM originating from various sources, including supernovae. These experiments can serve as effective tools for dark matter detection, complementing dedicated dark matter detectors. In the occurrence of the next nearby supernova--- possibly Betelgeuse---, its high flux would offer an exciting opportunity to constrain the dark-matter parameter space and study multi‑messenger astrophysics.

In the future, efforts can focus on refining models of dark-matter production from supernovae by improving simulations of BDM production and emission. Enhancements in background reduction will also be crucial for increasing the sensitivity of neutrino detectors to rare events, potentially including BDM signals. In addition to supernovae, other astrophysical sources such as active galactic nuclei, neutron star mergers, and cosmic rays can also serve as additional sources of BDM. All in all, neutrino detectors provide a promising avenue for testing various dark-matter models, particularly those involving low-mass dark matter and small couplings to Standard Model particles, making them attractive targets for further study and detection prospects.

\acknowledgments

The authors are grateful to Bhupal Dev, William DeRocco, Andrew Long, and Jong-Chul Park for insightful discussions. 
F.H., D.K., and K.S. thank the organizers of the Mitchell Conference in  May 2024 and 2025 at  Texas A\&M University for their hospitality and support during this project. 
They are also grateful to the organizers of the workshop of the Center for Theoretical Underground Physics and Related Areas (CETUP* - 2024 and 2025), the Institute for Underground Science at Sanford
Underground Research Facility (SURF), Lead, South Dakota 
for their hospitality and financial support. 

\bibliography{main.bbl}
\end{document}